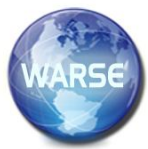

# Optimal Number of Cluster Heads in Wireless Sensors Networks Based on LEACH

Ghassan Samara [1], Munir Al-okour [2]
Department of Computer Science
Faculty of Information Technology
Zarqa University, Zarqa, Jordan

**ABSTRACT**

The Wireless Sensor Network (WSN) has been one of the leading research fields of wireless networks, particularly in recent year. Sensors are randomly positioned in the region, every node senses the surroundings and sends the data collected to the cluster head (CH), which aggregates and transmits obtained information to the Base Station (BS). A non-rechargeable battery is included on each WSN node. The sensor energy and network life extension of WSN are the most important considerations in the academia and industry. And thus, many routing protocols have been proposed to solve this issue, one of these is LEACH, the early protocol that introduced the clustering idea to extend the life of the WSN. LEACH is affected by the number of heads of the Clusters as it is randomly selected, and this has an impact on network lifetime, furthermore, the nodes are randomly joined in each cluster, this means that some of the cluster heads work more than others with fewer cluster nodes. In this paper, it is proposed an algorithm Maximum Optimal Number of Cluster Heads (MONCH) to identify the optimum cluster heads in WSN and to find which is the nearest one to BS and helps to integrate nodes with the most appropriate cluster. The results show improved energy consumption performance for the LEACH algorithm.

## 1. INTRODUCTION

The wireless sensor network consists of a set of enormous, cost-effective, wireless sensor nodes that are deployed at random in order to monitor a particular environment and send the collected information to base station BS, then BS [1], [2] aggregate the data and sends it to the users over the Internet.

WSNs are most commonly applied to monitor the environment, prediction of disasters, healthcare, intelligent housing, intelligent transport and other fields [3], [4]and [5]. Figure 1 shows the typical WSN cluster architecture [6] Clustering is a key mechanism for creating a scalable sensor system, The network is divided up into disjoint clusters, Each cluster consists of a head and joined nodes, where nodes scan the surrounding environment and transmit sensed data to cluster heads (CH) and which in turn aggregates the data [7], send aggregated data to the BS. Each node has rigid restrictions on the use of electricity, calculation and memory resources because the environment is hardly available and the battery replacements are not practical for a large number of nodes. The cluster heads used heavily by fusion data and their battery power draining quickly make it a challenge for researchers to develop and design algorithms for enhancing energy efficiency and enhancing WSN lifetime. [8].

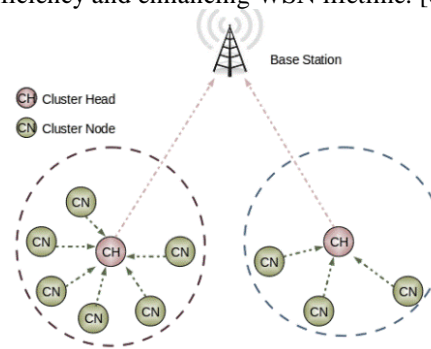

**Figure 1:** Clustering model

## 2. LEACH ALGORITHM

The LEACH idea mainly involves creating sensor node clusters based on the received signal strength and using the cluster heads as a rely point to the BS. Cluster head nodes are randomly chosen. The ordinary nodes called cluster members are joined by the principle of closeness to the appropriate cluster head nodes. Normal nodes sense data from the environment and directly send it to the head node of the cluster. The header nodes of the clusters get sensed data, aggregate it, eliminate the redundancy and forward it to the BS [9], [10].

LEACH is split into rounds in which each round contains two phases the setup phase and the steady-state phase.

In the setup phase, the cluster heads will be chosen by selecting a random number between 0 and 1, when the number is lower than a predefined threshold, the node will become a cluster head.

Then the node announces a cluster head advertisement message to its neighboring nodes, normal nodes subsequently send a join message to the CH that has the strongest signal received.

Each CH generates and distributes TDMA time slots to the members in its cluster after the network is arranged into clusters, also a CDMA code is selected to be used for sending sensed data to the BS.

The Steady-state Phase: Each sensor node starts sensing the environment and transmits the sensed data to its CH, CH node receives sensed data from all its group members, subsequently, aggregate the information and send them to BS. The network's status is restarted to setup and a fresh round is launched [11].

*Limitation Of Leach Algorithm*

Obviously, the LEACH has many advantages for WSNs in cluster organization, but it still has several shortcomings. LEACH determines cluster heads randomly without determining the maximum number of cluster heads.

LEACH formulate cluster heads randomly without determining which cluster head will formulate firstly.

LEACH does not determine the number of the join nodes to its cluster head.

For this limitations of LEACH algorithm, the study aims to improve LEACH algorithm to determine the optimal maximum number of cluster head and who will formulate firstly and the optimal number of the join nodes with its cluster head to develop clustering protocol, which has the primary role in the increasing age of the nodes and prolong the lifetime of the WSN.

## 3. RELATED WORK

Authors in [12] proposed LEACH-C is improved over LEACH using centralized clustering and has the same LEACH phases. LEACH-C varies in the setup phase the number of CHs is already determined at this point, where each node transmits its present residual energy level to the BS, nodes with energy more than the average value have the opportunity to be randomly chosen by the BS as clusters head. The BS subsequently broadcasts an advertising message to all network sensor nodes, this message includes CHs and Member Nodes identification, The results of the simulation show that LEACH-C has a clear improvement over LEACH due to the predetermined optimal number of cluster heads.

Authors proposed in [13] an algorithm to enhance cluster head selection technique in the setup phase, BS then examines whether or not the head of the cluster is balanced by comparing the real number of nodes to the average number of nodes, simulation results show an improved performance over the traditional LEACH.

In [14] authors proposed Kmedoids-LEACH (K-LEACH) protocol which uses the kmedoids clustering algorithm for uniform clustering, The cluster head (CH) is selected with euclidean distance and Maximum Residual Energy. The results of the simulation show that K-LEACH improves LEACH network lifetime.

LEACH-based SNCR is proposed in [15] This algorithm decreases the amount of singletons (cluster with only one node) and balances the energy consumption among the sensors, BS monitors the number of sensors per cluster, and assigns that node to the closest cluster when it finds a singleton cluster. The information then is sent to the sensors. The singleton cluster issue occurring in LEACH was resolved by this method without extra overhead control packets.

In [16], the authors proposed how to increase visibility of the network in aggregation techniques.

## 4. PROPOSED PROTOCOL MONCH

An important factor that affects the overall performance of a hierarchical routing protocol is an optimum number of cluster heads. Less CHs implies that a bigger region must apply to each CH. Less CHs implies that a bigger area must be covered by each CH so that more energy will be consumed. More CHs that implies, that every CH has to cover a smaller region, and more energy will be consumed. The size of the region, the number of nodes in the region, energy consumed by the head and node energy must be defined in some circumstances to determine optimal number covering the area. So, to determine the maximum optimal number of CHs (Kopt) use equation (1).

$$Kopt = \sqrt{\frac{Eh}{Ec} * 2 * pi * \frac{M}{dBS^2} * N} \quad ...(1)$$

Where dBS is obtained from equation 2 which the Euclidean distance for BS, M is the area of WSN and obtained from equation (3), Eh the energy consumed in the cluster head, which can get from equation 4.

$$dBS = \sqrt{BSx^2 + BSy^2} \quad ...(2)$$

$$M = \sqrt{YardLength^2 + YardWidth^2} \quad ...(3)$$

$$Eh = (ETX+EDA)*PacketLength + Efs*PacketLength \quad ...(4)$$

Where ETX is energy for transferring or receiving of a bit, EDA is data aggregation energy, and PacketLength is a fixed message length.

Ec the energy consumed in a node, which obtained from equation 5.

$$Ec = ETX*NodePacketLength + Efs*PacketLength \quad ...(5)$$

Where NodePacketLength is a constant node packet transfer.
Create first cluster head: The status information for each node is sent to the BS such as the (ID, position and rest of energy), with BS determining the closest head to the BS by Equation 6.

$$Dist(h,i) = D(i, BS) * Ei \quad ...(6)$$

Where D is the distance from nodes to BS which form the first cluster base on the closest distance.

Equation 7 is used to find the number of nodes in each head of the cluster (CN).

$$CN = \frac{N}{Kopt} \quad ...(7)$$

Where N is the number of nodes.

Equation 8 is used to locate the area where the first cluster will establish nodes.

$$DisHN(Xh, Yh) = \sum_{i=1}^{n} \sqrt{(Xh - Xi)^2 + (Yh - Yi)^{\wedge}2} \quad ...(8)$$

Where *DisHN(Xh, Yh)* distance node from cluster head, *(Xh,Yh)* position of cluster head, (Xi, Yi) location of node i, N is the number of remaining nodes.
The proposed algorithm is shown in Figure 2 below.

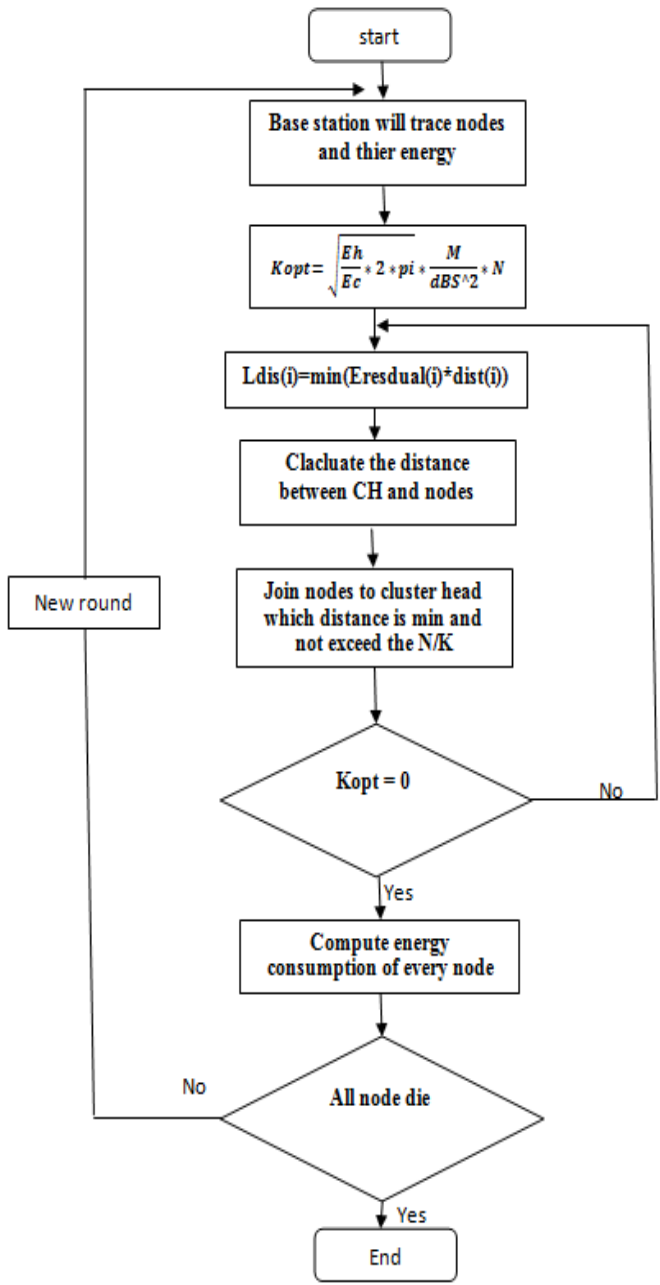

**Figure 2:** MONCH Algorithm Flowchart

## 5. ENERGY CONSUMPTION MODEL
The streamlined energy model mentioned in [9] is used for the energy consumption of radio communication. A distance d is used in equation 9 to convey a k-bit signal.

$$ETx(k) = \begin{cases} E_{elec} * k + E_{amp} * K * d^2 \text{ if } d < do \\ E_{elec} * k + E_{fs} * K * d^4 \text{ if } d > do \end{cases} \quad ...(9)$$

When Eelec radios are dissipated to operate the electronic transmitter or receiver, Eamp is the energy unit needed to amplify the transmitter in the free space; Efs is the unit energy necessary for the multipath fading transmitter amplifier, Do is the threshold of distance by equation 10.

$$do = \sqrt{\frac{E_{amp}}{E_{fs}}} \quad ...(10)$$

For the received k-bit message and distance d, equation 11 is used.

$$ERx(k) = E_{elec} * k \quad ...(11).$$

## 6. SIMULATION PARAMETERS AND EVALUATION METRICS
The performance metrics employed to measure network energy efficiency compared the LEACH algorithm to the proposed MONCH algorithm in terms of:
(1) Lifetime of each node.
(2) Residual energy of all node.
(3) Number of packets sent to the BS.

Table 1 below lists the simulation parameters.

**Table 1:** Simulation Parameters

| Parameter | Value | Parameter | Value |
|---|---|---|---|
| Network size | 100 nodes | YardLength | 100 m |
| YardWidth | 100 m | ETX | 50 nJ/bit |
| EDA (data aggregation energy) | 5 nJ/bit | Einit (Initial energy of node) | 2J |
| PacketLength | 6400 | NodePacketLength | 200 |
| Eamp (Radio amplifier energy) | 100 pJ/bit/m2 | Efs(Radio free space) | 0.013 pJ/bit/m4 |

In the network model, there are certain assumptions:
1- All sensor nodes in a square region are uniformly distributed.
After deployment, all sensors and the BS are stationary.
All nodes are battery powered and are not recharged.
The location of all nodes is known (e.g., by GPS).

## 7. SIMULATION RESULT
After both MONCH and LEACH algorithms are executed with the past parameters on MATLAB R2017b, Figure 3 shows the outcomes that demonstrate each node's lifetime.

The Y-axis shows the number of nodes and the X-axis shows the lifetime of the nodes in the network, Half the nodes died for MONCH at round 400 while at round 300 LEACH died. LEACH nodes did not actually last over 400 rounds, while MONCH nodes survived over 1400 rounds for the last nodes.

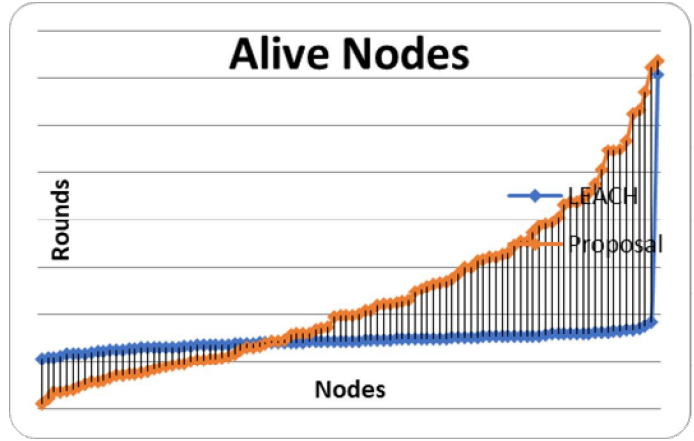

**Figure 3:** nodes alive

The power discharges on the basis of the executed activities such as sensing, information processing and communication, it is evident that the suggested MONCH algorithm results in greater energy efficiency than LEACH, Figure 4 demonstrates that, in round 421, LEACH mainly exhausted network energy while in round 1350, MONCH's energy nearly exhausted.

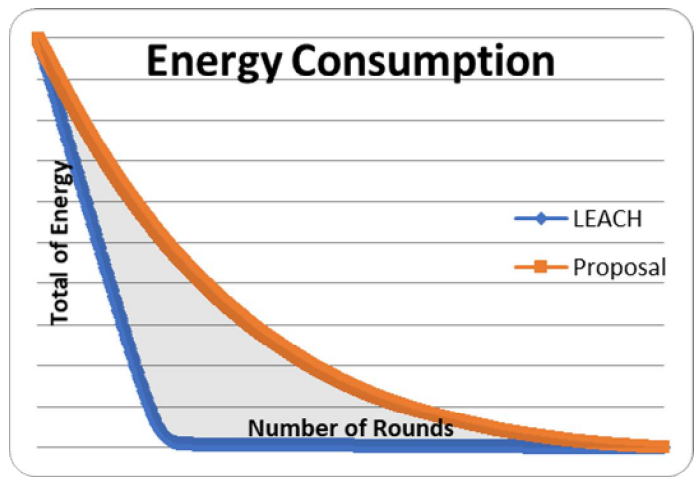

**Figure 4:** Energy consumption

Packets sent to BS in LEACH are given in Figure 5. In LEACH at round 421, 8352 is the majority of packets forwarded to BS, and at around 1477, BS is receiving 8346 Packets. While in MONCH The amount of packets forwarded to BS at round 421 is 13358, And at the round 1477 BS receives 20708 packets, This implies that the proposed MONCH's output extends the network lifetime to more than LEACH.

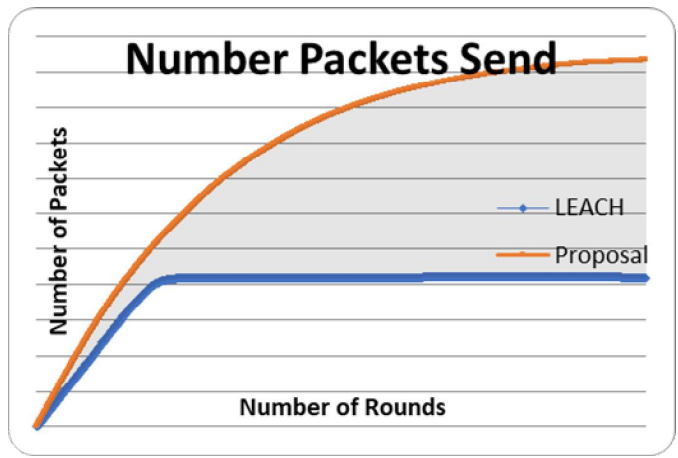

**Figure 5**: Number of Packet Sent to Base Station.

**ACKNOWLEDGEMENTS**
"This research is funded by the Deanship of Research in Zarqa University /Jordan."

## 8. CONCLUSION AND FUTURE WORKS
This paper proposed Maximum Optimal Number of Cluster Heads (MONCH) algorithm that improves the lifetime of WSN effectively, the proposed algorithm identified the optimum cluster heads in WSN and found which is the nearest one to BS and helped in integrating nodes with the most appropriate cluster. the results show an improved energy consumption performance over the LEACH algorithm

Ghassan Samara *et al.,* International Journal of Advanced Trends in Computer Science and Engineering, 9(1), January – February 2020, 891 – 8957. Heinzelman, W.R., Chandrakasan, A. and Balakrishnan, H., 2000, January. **Energy-efficient communication protocol for wireless microsensor networks**. In *Proceedings of the 33rd annual Hawaii international conference on system sciences* (pp. 10-pp). IEEE.
8. Vyas, P. and Chouhan, M., 2014. **Survey on clustering techniques in wireless sensor network**. *International Journal of Computer Science and Information Technologies*, *5*(5), pp.6614-661.
9. Dhawan, H. and Waraich, S., 2014. **A comparative study on LEACH routing protocol and its variants in wireless sensor networks: a survey**. *International Journal of Computer Applications*, *95*(8).
10. Dias, G.M., Bellalta, B. and Oechsner, S., 2016. **A survey about prediction-based data reduction in wireless sensor networks.** *ACM Computing Surveys (CSUR)*, *49*(3), pp.1-35.
11. Ma, Z., Li, G. and Gong, Q., 2016. **Improvement on leach-c protocol of wireless sensor network (leach-cc).** *International Journal of Future Generation Communication and Networking*, *9*(2), pp.183-192.
12. Xinhua, W. and Sheng, W., 2010, August. **Performance comparison of LEACH and LEACH-C protocols by NS2.** In *2010 Ninth International Symposium on Distributed Computing and Applications to Business, Engineering and Science* (pp. 254-258). IEEE. https://doi.org/10.1109/DCABES.2010.58
13. Saxena, P., Patra, P. and Kumar, N., 2015. **Energy Aware Approach in Leach Protocol for Balancing the Cluster Head in Setup Phase: An Application to Wireless Sensor Network.** *Journal of Information Assurance & Security*, *10*(1), pp.40-47.
14. Bakaraniya, P. and Mehta, S., 2013. **K-LEACH: An improved LEACH protocol for lifetime improvement in WSN.** *International Journal of Engineering Trends and Technology (IJETT)*, *4*(5), pp.1521-1526.
15. Bakaraniya, P. and Mehta, S., 2013. **K-LEACH: An improved LEACH protocol for lifetime improvement in WSN.** *International Journal of Engineering Trends and Technology (IJETT)*, *4*(5), pp.1521-1526.
16. Samara, G., Alsalihy, W.A.A. and Ramadass, S., 2011. **Increasing Network Visibility Using Coded Repetition Beacon Piggybacking**. *World Applied Sciences Journal*, 13(1), pp.100-108.
895